
\magnification=1100
\hsize=14.8 true cm
\vsize=22.5 true cm
\hoffset=1.3 true cm
\voffset=0.1 true cm
\baselineskip=16 pt

\font\large=cmbx10 scaled\magstep1
\vbox{\baselineskip=11 pt
\line{\hfill May 1995 } }
\vskip .9 true cm

\centerline{\large
Universality of the Tearing Phase in Matrix Models }

\vskip 1 true cm
\noindent
\hskip 5.2truecm {\bf Giovanni M.Cicuta} \footnote*
{Cicuta@PARMA.INFN.IT}
\vskip .3truecm
\centerline{
Dipartimento di Fisica, Universit\`a di Parma}
\centerline{and INFN, Gruppo collegato di Parma}
\centerline{Viale delle Scienze, 43100 Parma}
\vskip .5 true cm
\noindent
\hskip 5 true cm {\bf Luca Molinari} \footnote*{Molinari@MILANO.INFN.IT}
\vskip .3 true cm
\centerline{Dipartimento di Fisica, Universit\`a di Milano}
\centerline{ and INFN, Sezione di Milano}
\centerline{Via Celoria 16, 20133 Milano}
\vskip .5 true cm
\noindent
\hskip 4.8 true cm {\bf Sebastiano Stramaglia} \footnote*{IRMASS12@AREA.BA.
CNR.IT}
\vskip .3 true cm
\centerline{
Istituto per Ricerche di Matematica Applicata, C.N.R.}
\centerline{Via Amendola 122I, 70100 Bari}
\vskip .8 true cm
{\bf Abstract.}  The spontaneous symmetry breaking associated to the tearing
of a random surface, where large dynamical holes fill the surface, was
recently analized obtaining a non-universal critical exponent on a border
phase. Here the issue of universality is explained by an independent
analysis. The one hole sector of the model is useful to manifest the
origin of the (limited) non-universal behaviour, that is the existence
of two inequivalent critical points.

\vfill \eject

Some years ago, Kazakov analyzed an interesting matrix model describing random
surfaces with dynamical holes [1]. He showed that the continuum limit of the
model
has three phases: a "perturbative" one where small holes do not alter the
geometric properties of the random surface; a "tearing phase", where the
surface is formed by thin strips surrounding large holes of diverging average
length, and a third phase separating the above two where both holes and strips
are large and competing.
\par
The model was analyzed mainly as a solvable model of open strings embedded in
zero or in one dimension [1,2,3]. Matrix models in reduced dimensions provide
interesting statistical models and are suggestive of critical behaviours that
may occur in the non perturbative analysis of quantum field theory in more
realistic dimension of space time.
\par
Due to the very interesting features of the tearing transition, with the aim of
testing the universality of its critical exponents, a matrix model very similar
both to Kazakov's model [1] and to the $O(n)$ matrix model [4,5] was recently
analyzed [6].
The qualitative description of the critical behaviour turned to be equivalent,
with the same critical exponents in the two critical phases (small holes
phase or perturbative phase, and the tearing phase) but a different
one on the border phase.
This discrepancy was unexpected and it deserves deeper understanding, provided
by the present letter.
\par
In the first part of this work we study the model by the method of orthogonal
polynomials and find the scaling laws which characterize the scaling behaviour
of the holes on the random surface in our model. In the second part we restrict
the model to the "one -hole" sector: we investigate the case of a "static"
loop interacting with the random surface. A first-order transition is found,
corresponding to the hole filling the surface. This first-order transition is
the memory of the tearing phase in the one-hole sector.
\par
The partition function of the model is
$$
Z_N (L,g,z) = \int {\cal D} M \; {\rm exp} \{ -N{\rm Tr} [ V(M) + L \,
 \log(1 -2zM) ] \}  \eqno (1)
$$
where $M$ is an hermitian $N\times N$ matrix, $g>0$,$z>0$ and the potential is
$$V(M)={1\over 2}M^2 + {g\over 3}M^3 \eqno (2)$$
\par
We refer the reader to ref.[6] for the notations and an analysis of the
relations between this model and Kazakov's model as well as the $O(n)$
model.
\par
Let us introduce $\lambda =g^2$, $\gamma =g^2L$ and $\mu ={2z\over g}$. The
fugacity of the number of holes on the surface is $\gamma$, while $\mu$ turns
out to be the effective fugacity of the total perimeter of the holes.
After a rescaling $\phi =gM$ the partition function may be written as
$$Z=\int d\phi \exp{ -{\rm tr}{N\over \lambda}\left( {1\over 2}\phi^2
+ {1\over 3}\phi^3 + \gamma {\ln}(1-\mu\phi)\right)}=\int d\phi \exp{-{\rm
tr}{N\over \lambda}{\cal V}(\phi )} \eqno (3)$$
\par
The matrix model can be studied by the standard method of orthogonal
polynomials [7]. Let us introduce the set $\langle\varphi |n\rangle
=P_n (\varphi )$ of
polynomials
orthonormal with respect to the measure $d\mu=d\varphi e^{-{N\over
\lambda}{\cal V}(\varphi )}$:
$$\langle m|n\rangle =\int d\mu P_m(\varphi ) P_n(\varphi ) =\delta_{mn}
\eqno (4)$$
\par
The coordinate operator $\hat \varphi :g(\varphi )\to \varphi g
(\varphi )$ has the following matrix elements:
$$\langle m|\hat \varphi |n\rangle =\sqrt{R_m}\delta_{m,n+1} + S_n \delta_{m,n}
+ \sqrt{R_n}\delta_{m,n-1} \eqno (5)$$
Then
$$Z=N!C^N \prod_{i=1}^{N-1} R_{i}^{N-i} \eqno (6)$$
where the constant $C$ is the normalization of the measure: $C=\int d\mu$.
\par
The coefficients $R_n$ and $S_n$ are determined by the "equations of motion":
$$\langle n|{\cal V}'(\hat \varphi )|n\rangle = 0 \eqno (7a)$$
$$\langle n-1|{\cal V}'(\hat \varphi )|n\rangle ={n\lambda \over
{N\sqrt{R_n}}} \eqno (7b)$$
which for our potential have the form:
$$0=S_n+S_n^2+R_{n+1}+R_{n}-\mu\gamma\langle n|{1\over {1-\mu \hat
\varphi}}|n\rangle \eqno (8a)$$
$${n\lambda \over {N\sqrt{R_n}}}=\sqrt{R_n}(1+S_n+S_{n-1})-\mu\gamma\langle
n-1|{1\over {1-\mu\hat \varphi}}|n\rangle \eqno (8b)$$
\par
The operator ${(1-\mu\hat \varphi )}^{-1}$ is the resolvent of a
random motion on the lattice {\bf N}.
In order to perform the planar limit $N\to\infty$ it is convenient to introduce
the conjugate operators $\hat l$ and $\hat \theta$ [8], defined by
$${\hat l} |n\rangle = {n\over N} |n\rangle\;,\;e^{\pm i\hat \theta}|n\rangle =
|n\pm 1\rangle \eqno (9)$$
The operator $\hat \varphi$ can be expressed as
$${\hat \varphi} =\sqrt{R({\hat l})} e^{i\hat \theta} +S({\hat l})+e^{-i\hat
\theta}\sqrt{R({\hat l})} \eqno (10)$$
and in the $\theta -basis$, $|\theta\rangle ={1\over
\sqrt{2\pi}}\sum_{n}e^{in\theta}|n\rangle$, $\hat l$ acts as a derivative.
\par
In the large $N$ limit $\hat l$ commutes with $\hat \theta$ and can be taken
equal to the identity. The operator $\hat \varphi$ simplifies to:
$$\hat \varphi =2 \sqrt{R}cos\theta +S \eqno (11)$$
having assumed the limits
$$R=\lim_{n\to\infty} R_n\;,\;S=\lim_{n\to\infty} S_n$$
In the planar limit we have
$$ \langle n|{(1-\mu\hat\varphi )}^{-1}|n\rangle \to \int_{0}^{2\pi}
{d\theta\over 2\pi} {1\over 1-\mu\varphi (\theta)}={1\over \sqrt{{(1-\mu
S)}^2-4\mu^2 R}} \eqno (12)$$
and
$$\langle n-1|{(1-\mu\hat\varphi )}^{-1}|n\rangle \to
\int_{0}^{2\pi}{d\theta\over 2\pi}{e^{i\theta}\over {1-\mu\varphi (\theta )}}=
{1\over 2\mu\sqrt{R}}\left( {{1-\mu S}\over \sqrt{{(1-\mu S)}^2-4\mu^2R}}-1
\right) \eqno (13)$$
The equations of motion in the planar limit read
$$0=S+S^2+2R-\mu\gamma {G^{-1}(R,S,\mu )}
\eqno (14a)$$
$$\lambda =R+2RS +{\gamma\over 2}\big [1-(1-\mu S) G^{-1}(R,S,\mu )
\big ] \eqno (14b)$$
where we denote $G(R,S,\mu )=\sqrt{{(1-\mu S)}^2 -4\mu^2 R}$;
eqs.(14) correspond to eqs.(4.8) in ref.[6] with the identifications
$$S=\sigma\;,\;R={\delta^2\over 4}\; ,\; \mu ={1\over \tau}$$
\par
The above equations provide $S=S(\mu ,\gamma ,\lambda )$ and $R=R(\mu ,\gamma ,
\lambda )$. The continuum limit of the system corresponds to a critical
surface $f(\mu ,
\gamma ,\lambda )=0$ in the three-dimensional parameter space spanned by the
variables $\mu$, $\gamma$ and $\lambda$.
We can ensure critical behaviour by imposing the following scaling:
$$S=S_0 +S_1 a\;,\;R=R_0 +R_1 a$$
$$\lambda =\lambda_0 +\Lambda a^{\ell}\;,\; \gamma  =\gamma_0 +\Gamma a^{k}
\eqno (15)$$
with $\ell >1$ and $k>1$, where $a$ is a cut-off vanishing in the continuum
limit;
indeed eqs.(15) imply ${\partial S\over \partial \lambda}=\infty =
{\partial R\over \partial \lambda}$.
\par
The condition $G(\mu ,S_0,R_0)\ne 0$ characterizes the perturbative phase.
Inserting the scaling laws (15) in eqs.(14) and requiring non-trivial solutions
for $S_1$ and $R_1$ leads to the equation:
$$4R_0{(1+3S_0-{\mu}^{-1})}^2={[(1+2S_0)({\mu}^{-1}-S_0)-S_0(1+S_0)-6R_0]}^2
\eqno (16)$$
fully equivalent to the critical equation (4.9) in ref.[6]. As in Kazakov's
model [1], the analysis of the critical behaviour is simplified by considering
the values of $\gamma$ with $\gamma_0 =0$. The critical values
are then $S_0={-3+\sqrt{3}\over 6}$, $R_0={1\over 12}$, $\lambda_0={1\over
12\sqrt{3}}$;
the consistent value for the exponents $\ell$ and $k$ is $2$. This is the
"small holes" phase [1,6].
\par
Let us now consider the non-perturbative phase.
\par
When $G(\mu ,R_0,S_0)$ tends to zero as $\gamma$ vanishes a new critical
behaviour arises: the phenomenon of spontaneous tearing discussed in ref.[1,6].
Inserting the scaling laws (15) ,with $\gamma_0=0$ and the condition
$G(\mu ,R_0,S_0)=0$, in eqs.(14) (we observe that in the non-perturbative phase
$\ell$ is not supposed to be greater than one because criticality is ensured by
the vanishing of $G$), we obtain:
$$R_0(\mu )=-{1\over 2}S_0(S_0+1)\;,\;S_0(\mu )={1\over 3\mu}\left( 1-\mu
+\sqrt{{(1-\mu)}^2-3}\right)
$$
$$\lambda_0(\mu )=-{1\over 2}S_0(1+S_0)(1+2S_0) \eqno (17)$$
and the equation
$$2\mu^2\Lambda=(\mu+3\mu S_0-1)[2R_1-{S_1\over \mu}(1-\mu S_0)] \eqno (18)$$
The square root in (17) implies $\mu\ge\mu_c =(1+\sqrt{3})$ for the
non-perturbative phase, and the consistent values for $\ell$ and $k$ are
respectively $1$ and $3/2$.
\par
The case $\mu =\mu_c$ (critical tearing) has to be investigated separately,
since eq.(18) implies
$\Lambda =0$ in this limit.
Note that eqs.(14) may be rewritten as
$$R={\lambda -{\gamma\over 2}+{S\over 2\mu}(1+S)(1-\mu S)\over
1+3S-\mu^{-1}}\eqno (19a)$$
$$0=S(1+S)(1+2S)+2\lambda -\gamma -\mu\gamma (1+3S-\mu^{-1})G^{-1} \eqno
(19b)$$
with $1+3S_0-\mu_{c}^{-1}=0$.
It is straightforward to check that the scaling law compatible with eqs.(19)
when $\mu\to\mu_c$ is
$$R=R_0 +R_1 a\;,\;S=S_0+S_1 a\;,\;\mu =\mu_c -M a$$
$$\lambda =\lambda_0+\Lambda a^{3/2}\;,\;\gamma =\Gamma a^{3/2} \eqno (20)$$
to be compared with the corresponding law in Kazakov's model [2]:
$$R=R_0 +R_1 a\;,\;\mu =\mu_c -M a$$
$$\lambda =\lambda_0 +\Lambda a^2\;,\;\gamma =\Gamma a^{5/2} \eqno (21)$$
\par
It follows that in our model the dynamical holes exhibit, in the intermediate
phase, a different scaling behaviour with respect to Kazakov's model. Indeed
the typical area of the surface diverges at criticality as ${1\over \lambda_c
-\lambda}$, while the total perimeter of the holes on the surface diverges
as ${1\over \mu_c -\mu}$. Then in the intermediate phase (critical tearing)
the scaling laws (20)
imply for our
model that the "length" of the holes scales as the area to the power of $2/3$,
while eqs.(21) imply for Kazakov's model that in the intermediate phase
the length of the holes scales as the square root of the area.
\par
It is interesting to observe that if we defined our model with the
potential
$$V_1(M)={1\over 2}M^2 - {g\over 3}M^3\;,\;g>0 \eqno (22)$$
instead of (2), then the equations corresponding to (19) would be:
$$R={\lambda -{\gamma\over 2}+{S\over 2\mu}(1-S)(1-\mu S)\over 1-3S+\mu^{-1}}
\eqno (23a)$$
$$0=S(1-S)(1-2S)-2\lambda +\gamma -\mu\gamma (1-3S+\mu^{-1})G^{-1} \eqno
(23b)$$
with the critical values $S_0={3-\sqrt{3}\over 6}$ ,$R_0={1\over 12}$ and
$\mu_c =3-\sqrt{3}$ the positive solution of the equation
$${(1-\mu_c S_0)}^2 -4\mu_c^2 R_0 =0$$
In this case $1-3S_0 +\mu_c^{-1} \ne 0$ and eqs.(23) admit a scaling law
completely analogous to (21) implying the same scaling behaviour for the holes
as in Kazakov's model even in the intermediate phase.
\par
Let us explain this point. The one matrix model
$$V(M)= {1\over 2}M^2 + {g\over 3}M^3$$
is invariant under $g\to -g$ and $M\to -M$, so it has two critical points,
$g^{*}$ and $-g^{*}$. The two critical points are equivalent for the pure cubic
model and they both describe pure gravity. When the random surface is coupled
to the holes the two critical points are no more equivalent: if the surface
reaches the continuum limit by sending $g$ to $g^{*}$ then the holes always
have
the same scaling behaviour as in Kazakov's model, while sending $g$ to $-g^{*}$
the holes have a different scaling behaviour in the intermediate phase.
\par
The "anomalous" scaling behaviour of the dynamical holes in the intermediate
phase is connected with the following feature of our model in the one-hole
sector:
the absence of the dilute phase for the single static hole interacting with the
random surface.
\par
The one-hole sector is obtained, as explained in ref.[6], considering the
formal Taylor expansion in $L$ of the free energy of model (1)
$$E(L,g,z)=-\lim_{N\to\infty}{1\over N^2}\ln Z_{N}=\sum_{k=0}^{\infty}L^k E_{k}
(g,z) \eqno (24)$$
The term $E_1$ is the generator of planar connected graphs with one hole.
In terms of the density of eigenvalues $\rho_3 (\lambda )$ of the pure one
matrix cubic model [9] $E_1$ is given by the following integral:
$$E_1(g,z)=\int_{a_0}^{b_0} d\lambda \rho_3(\lambda ) \ln{(1-2z\lambda)} \eqno
(25)$$
The singularity $g_c(\tau )$ of $E_1$ (where $\tau ={g\over 2z}$) yields the
density of free energy in the thermodynamic limit [6]:
$$f=\ln{g_c (\tau )}$$
by means of which we can evaluate the average length of the perimeter of the
hole per unit area ${\cal L}={\partial f\over \partial \ln{\tau }}$.
\par
The density of length of the hole ${\cal L}$ plays here the role of an order
parameter: ${\cal L}=0$ corresponds to a "confined" polymer having a finite
perimeter, ${\cal L}\ne 0$ corresponds to a polymer with infinite length which
is dense on the surface (indeed the hole boundary is a fractal).
\par
The explicit expression of $E_1$ has been evaluated in [6]. Its singularity may
arise from the singularity of $\rho_3$ with respect to $g$ or from the
vanishing of the argument of the log in eq.(25) at $\lambda =b_0$, i.e.
condition ${1\over 2z}=b_0$.
\par
In the range $\tau >\tau_c ={1\over 2}(\sqrt{3}-1)$ the singularity of $E_1$ is
given by the singularity of $\rho_0$ and ${\cal L}=0$:
$$g_c^2={1\over 12\sqrt{3}}\;,\;\tau>\tau_c \eqno (26)$$
In the range $0<\tau <\tau_c$ the singularity is due to the condition ${1\over
2z}=b_0$ and the parametric expression of $g_c$ is
$$2g_c^2 +\sigma (1+\sigma )(1+2\sigma )=0$$
$$\tau =\sigma +\sqrt{-2\sigma (1+\sigma )}\;,\;0<\tau<\tau_c \eqno (27)$$
which imply ${\cal L}\ne 0$, i.e. the hole is dense on the surface.
In fig.(1a) $g_c$ versus $\tau$ is plotted. We see that $g_c (\tau )$ is
continous at $\tau_c$ but its first derivate (proportional to ${\cal L}$) is
not. Hence a first-order
transition occurs with the absence of the dilute phase for the polymer [5].
\par
The one-hole sector in Kazakov's model is defined by
$$E_1 = \int_{-a}^{a}d\lambda \rho_4 (\lambda )\ln{(1-z^2\lambda^2)} \eqno (28)
$$
with $\rho_4(\lambda )$ being the density for the pure quartic model [9].
By setting $ \tau={g\over z^2}$ , one easily finds
$$g_c(\tau )={\tau\over 4}-{3 \tau ^2 \over 16}\;,\;0<\tau <2/3$$
$$g_c(\tau )={1\over 12}\;,\;\tau>2/3  \eqno (29)$$
The phase transition is second order. In fig.(1b) the critical coupling of
Kazakov's model in the one-hole sector is plotted versus the fugacity. In this
case a dilute phase for the polymer is found, corresponding to the critical
fugacity, where the length of the polymer scales as the square root of the area
of the surface.
\par
Let us now compare our results for the single hole with the analysis
of self avoiding walks on random surface [5]. The case in which two random
walks tied together at their ends live on a random trivalent lattice is
equivalent to our $E_1$ model with the difference that the two SAWs form a
loop but not one hole. In fig.(1c) the critical coupling versus the fugacity is
plotted in this case, showing that a second order transition occurs. The
transition point corresponds to the dilute phase. It is interesting to observe
that the model with $V_1(M)$, eq.(22), instead of (2) in model (1) has in the
one-hole sector a critical curve $g_c(\tau )$ for the single hole which is
exactly the same as the one plotted in fig.(1c) for the
two SAWs. Conversely changing the sign of the coupling constant in the two SAWs
model yields exactly the critical curve in fig.(1a) and implies
the absence of the dilute phase.
\par
These results seem to suggest the following relation between a model of
dynamical holes on a random surface and the corresponding one-hole sector:
the transition in the one-hole sector is second-order if and only if
the dynamical holes have at critical tearing the "standard" scaling behaviour
(with the length of the holes scaling as the square root of the area of the
surface).
\par
Let us summarize the main results of this letter:
\par
\item{(1)} We exhibit the scaling behaviours proper to the continuum limit
for the three critical phases and we confirm the critical exponents found in
[6].
\item{(2)} The existence of two inequivalent critical points for the model of
self avoiding walks on random surfaces is here shown. The two points occur for
opposite values of the cubic coupling. They correspond, in models with
dynamical loops, to two inequivalent critical lines. One was described in the
paper [6], the other, also discussed here, yields the same exponents as the
critical line in Kazakov 's model.
\par

\vskip 4 truecm

{\bf Figure Captions}
\par
{\bf Fig.1} Critical coupling $g_c( \tau )$ of the one-hole sector ($\tau$ is
the inverse of the fugacity of the length of the polymer) for:
\par
\item{(a)} Our model, see eqs.(26,27).
\item{(b)} Kazakov's model, see eq.(29).
\item{(c)} Self-avoiding-walks, ref.[5], or our model with potential $V_1(M)$,
eq.(22).
\par

\vskip 4 truecm

{\bf REFERENCES}
\par\noindent
[1] V.A.Kazakov, Phys.Lett.{\bf B237}(1990) 212.
\par\noindent
[2] I.K.Kostov, Phys.Lett.{\bf B238}(1990) 181.
\par\noindent
[3] Z.Yang, Phys.Lett.{\bf B257}(1991) 40.
\par\noindent
[4] I.K.Kostov, Mod.Phys.Lett.{\bf A4}(1989) 217.
\par\noindent
[5] B.Duplantier, I.Kostov, Nucl.Phys.{\bf B340}(1990)491
\par\noindent
[6] G.M.Cicuta,L.Molinari,E.Montaldi,S.Stramaglia, preprint hep-th/9412080

and revised edition in May 1995.
\par\noindent
[7] C.Itzykson, J.B.Zuber, J.Math.Phys.{\bf 21}(1980)411.
\par\noindent
[8] O.Alvarez, P.Windey, Nucl.Phys.{\bf B348}(1991) 490.
\par\noindent
[9] E.Brezin, C.Itzykson, G.Parisi, J.B.Zuber, Comm.Math.Phys.{\bf 59}(1978)35

\vfill \eject
\bye